\documentclass[a4paper,11pt]{article}
\usepackage{epsfig} \usepackage{amsmath} \usepackage{amsfonts}
\usepackage{proteins}
\usepackage{threeparttable}
\begin{document}
\date{\today}
\begin{sloppy}
\begin{center}
{\Large \bf {On the performance of combined dichotomic predictors of natively unfolded proteins.}}
\end{center}
\begin{center}
{\large \bf {Antonio Deiana$^{1,2}$, Andrea Giansanti$^{1,2,3*}$} }
\end{center}

{\it $^{1}$ Physics Department, La Sapienza University of Rome, P.le A. Moro, 2, 00185, Rome, Italy}\\
{\it $^{2}$ Interdepartmental Research Centre for Models and Information Analysis in Biomedical Systems (CISB), La Sapienza University of Rome, C.so Vittorio Emanuele II 244, 00186, Rome, Italy}\\
{\it $^{3}$ INFN, Sezione di Roma, P.le A. Moro 2, 00185, Rome, Italy} \\

\noindent
$^*$ Correspondence to: Andrea.Giansanti@roma1.infn.it\\

\newpage
%
%
\begin{center}
{\bf {\it ABSTRACT}}
\end{center}

In this work we revisit ab-initio methods to identify natively unfolded proteins. Single predictors and combined score indexes are considered and their performance is critically evaluated against other methods already present in the literature. We consider mean packing ($<P>$), mean pairwise energy (<$E_c$>) and a new index of folding status, based on VSL2 ($gVSL2$), a predictor of single disordered amino acids. We use a dataset made of 743 folded proteins and 81 natively unfolded proteins. Individual use of these predictors has a performance comparable or even better than other proposed methods: gVSL2 reaches a sensitivity of 0.81, a specificity  of 0.89 and a level of false predictions of 0.11. Moreover, the performance of these single predictors is not significantly improved if used in combination. In many cases different predictors differently classified as folded or unfolded the same amino acid sequences. We introduce a strictly unanimous combination score $S_{SU}$ that combines mean packing, mean pairwise energy and $gVSL2$. $S_{SU}$ leaves undecided sequences differently classified by two single predictors. By applying the single indexes on dataset purged from the proteins left unpredicted  by $S_{SU}$, their performance significantly increases, indicating that unclassified proteins by $S_{SU}$ are mainly false predictions. Amino acid composition is the main determinant considered by predictors of natively unfolded proteins, therefore the classification of proteins with amino acid composition compatible with both folded and unfolded sequences is quite challenging. For this reason, if we exclude these proteins from the dataset, the performance of the folding indexes increases. The percentage of proteins predicted as natively unfolded by $S_{SU}$ in the three kingdoms are: 4.1\% for Bacteria, 1.0\% for Archaea and 20.0\% for Eukarya; comparable, but not coincident with similar previous determinations. Evidence is given of a scaling law relating the number of natively unfolded proteins with the total number of proteins in a genome; a first estimate of the critical exponent is 1.95 $\pm$ 0.21. 

\newpage
\begin{center}
{\bf {BACKGROUND}}
\end{center}
In the past few years it has been discovered that several proteins, in physiological conditions, lack a well defined tertiary structure, existing as an ensemble of flexible conformations. These proteins, denoted in the literature as \textit{natively unfolded} or \textit{intrinsically disordered}, are characterized, microscopically, by an high atomic diffusivity all along their sequence. Nevertheless they are involved in important cellular functions, like signalling, targeting or DNA binding \cite{Wright1999,Dunker2001,Demchenko2001,Uversky2002a,Tompa2002,Fink2005,Uversky2005}; their existence shifts the structure-function paradigm, that regards the tertiary structure of a protein as necessary for its biological function \cite{Creighton1993}. It has been suggested that natively unfolded proteins may also play critical roles in the development of cancer \cite{Iakoucheva2002}; moreover, the absence of a rigid structure allows them to bind different targets with high specificity and low affinity, suggesting that they are hubs in protein interaction networks \cite{Dunker2005,Heynes2006,Dosztanyi2006}. Unstructured regions may be present also in folded proteins. A specific local flexibility in these \textit{partially unfolded} proteins might play a dynamical role in modulating their interactions with other macromolecules. \\
In this work we critically review sequence-only, \textit{ab-initio}, methods to identify natively unfolded proteins. Computational approaches aimed at identifying unstructured regions in proteins are very useful, since the experimental characterization of these regions is flawed by a certain ambiguity, due to the several techniques available, that often give conflicting views on the same protein \cite{Daughdrill1999,Rose2002}. In particular, predictors of natively unfolded proteins may be useful to fastly screen datasets of amino acid sequences, looking for those that have a high tendency to remain unfolded; and this is the main application that we have in mind in this work. \\
Several methods have been proposed to predict unstructured segments in proteins \cite{Romero1997a,Romero1997b,Romero2000,Romero2001,Dunker2002b,Obradovic2003,Linding2003a,Linding2003b,Vucetic2003,Ward2004,Coeytaux2005,Dosztanyi2005a,Peng2006,Galzitskaya2006a}. These methods aim at identifying disordered amino acids, i.e. residues for which it is hard to determine experimentally, using X-ray cristallography or NMR spectroscopy, the average positions of their atoms \cite{Bordoli2007}. Predictors of disordered amino acids are useful to find unstructured regions in partially unfolded proteins, but they do not highlight immediately whether a protein globally folds or not. Besides, unfolded segments may have a wealth of different static and dynamic properties, but each predictor is generally focussed on just one specific characteristic, therefore it seems wise to combine the information from different indicators to obtain robust predictions \cite{Ferron2006,Dosztanyi2007}. Other methods aim at predicting whether a protein is globally natively unfolded. Several physico-chemical properties have been recognized as useful indicators, but they have been used differently by authors \cite{Uversky2000,Oldfield2005,Galzitskaya2006a,Dosztanyi2005a,Meszaros2007,Shimizu2007}; moreover the proposed methods have been tested on various datasets, so it is not easy to make comparisons, searching for an optimal approach. 
Within the present study we revisited and optimized various methods of predicting natively unfolded proteins. Two predictors are based on mean packing \cite{Galzitskaya2006a} and mean pairwise energy \cite{Thomas1996,Dosztanyi2005a}. A third index, called here gVSL2, is derived from VSL2 \cite{Obradovic2005,Peng2006}, a predictor of disordered amino acids that excellently performed in the recent CASP7 experiment \cite{Bordoli2007}. We retuned the parameters of the predictors, increasing the performance with respect to the original settings indicated in the literature. With our settings we discriminated folded proteins from natively unfolded ones with sensitivity up to 0.74 and a level of false predictions below 0.11. We then studied the possibility of enhancing the performance of single folding indexes by combining them into scores. We considered both the unanimous score $S_U$ and the voting score $S_V$ proposed by Oldfield $et$ $al.$ \cite{Oldfield2005}. We introduced also a score $S_0$ that requires consensus among the majority of folding indexes (see Methods for details) However, we did not succeeded in enhancing significantly the performance of single folding indexes. Interestingly, we observed an increasing of the performance with respect to single folding indexes by introducing a strictly unanimous score $S_{SU}$ that requires unanimous consensus among the various indexes of fold to classify a protein in one of the two folding classes, and that left unclassified proteins differently classified by at least two indexes. This score exhibited a sensitivity of 0.82, a level of false predictions of 0.05 and it left unclassified about 10\% of the proteins in the test set. It is important to note that the predictors of natively unfolded proteins currently described in the literature and reviewed in this paper classify a protein as folded or not by analysing mainly amino acidic composition of protein sequences (see Results and Discussion below and also \cite{Szilagyi2008}). In a nutshell, predictors of natively unfolded proteins are learning machine trained at recognizing sequences with amino acidic composition similar to that of the folded and natively unfolded proteins in a curated training set. Our results indicate that in a generic dataset there are proteins with amino acidic composition typical of folded and natively unfolded sequences; these proteins have a strong folding signature in their sequence and therefore they are generally classified correctly by all the predictors of natively unfolded proteins. However, in a dataset there are also proteins with amino acidic composition different from that of folded and natively unfolded sequences. These proteins do not have a well-defined folding signature in their sequence and they are often misclassified by single folding indexes. A score that does not classify them therefore reduces the number of false predictions and exhibits higher performance. Proteins with a weak folding signature cannot be analysed with the predictors of natively unfolded proteins currently described in the literature, but differently approaches must be used. \\
We applied our indexes to evaluate the frequency of natively unfolded proteins present in various genomes, obtaining results consistent with those reported by Ward $et$ $al.$ using DISOPRED2 \cite{Ward2004}. Since our approach is quite different from theirs, we think that it is a valid alternative. Finally, we observed a significant correlation, using our approach, between the number of predicted disordered proteins and the number of proteins in genomes of Bacteria, Archaea and Eukarya and we determined a scaling law, of possible fundamental significance, to be validated by further studies. \\

\noindent

\vspace{1cm}
\begin{center}
{\bf {RESULTS AND DISCUSSION}}
\end{center}

\noindent
{\bf Mean packing, mean contact energy and $gVSL2$} \\

In this work we revisited several predictors of globally unfolded proteins. We considered $HQ$, mean packing $<P>$, mean pairwise energy $<E_C>$ and $gVSL2$. $HQ$ is an our implementation of the method by Uversky and co-workers \cite{Uversky2000, Prilusky2005}. The mean packing of a protein is the arithmetic mean of the packing values of its amino acids. The packing of an amino acid is defined as the average number of its close residues, i.e. residues within a distance of 8 \AA, computed on a large set of structured proteins \cite{Galzitskaya2006a}. We considered as natively unfolded protein sequences with mean packing below 20.54. The mean pairwise energy of a protein is the arithmetic mean of the contact energy values of its amino acids. It was computed following the protocol implemented in the IUPred code \cite{Dosztanyi2005a}. We considered as natively unfolded protein sequences with mean pairwise energy above -0.37 arbitrary energy unit (a.e.u.). $gVSL2$ is an index derived from the disorder predictor VSL2 \cite{Obradovic2005,Peng2006}; specifically, gVSL2 is the arithmetic mean of the VSL2 scores, over the sequence. Details on the implementation we used are reported in the Methods section. We tested the performance of HQ, mean packing, mean pairwise energy and gVSL2 on a test set made of 743 folded and 81 natively unfolded proteins. The results are reported in table \ref{tab:Perf_idxs}. As we can see, HQ has, relatively, the worst performance. Mean packing and mean pairwise energy show similar performance, whereas gVSL2 has a comparatively higher sensitivity, but also a higher level of false predictions. \\
Mean packing and mean pairwise energy have been used previously to predict whether a protein is natively unfolded or not. Mean packing has been used by Galzitskaya and co-workers \cite{Galzitskaya2006a}. They used a sliding window restricted to just one amino acid and a threshold at 20.73. Using their setting on our own test set, the sensitivity arose from 0.74 to 0.83. However, the level of false predictions also grew, from 0.07 to 0.19. This suggests that, using the approach in \cite{Galzitskaya2006a}, one could overestimate the number of natively unfolded proteins present in the genome of a given organism. Dosztanyi $et$ $al.$ consider as disordered amino acids with contact energy value above -0.2 a.e.u. \cite{Dosztanyi2005a}; recently, this threshold has been used to effectively discriminate folded proteins from natively unfolded ones, in a peculiar set of protein complexes \cite{Meszaros2007}. Using the discriminative threshold of -0.2 a.e.u. on our test set, sensitivity dropped from 0.74 to 0.54. This result shows that the effectiveness of discriminative threshold, using single predictors, strongly depends on the chosen test set. However, our settings seems to be more robust that the those previously indicated in the literature. \\

\noindent
{\bf Shuffling-invariance of the folding indexes and their dependence from amino acidic composition} \\
We checked that the indexes of fold we analysed here are invariant under shuffling of the amino acids in the sequences (changes limited to a few percent). 
This shuffling invariance of the indexes suggests some considerations. There is a large consensus that the tertiary structure of a protein is stabilized by hydrophobic effects and Van der Waals interactions, not so sensible to the detailed geometry of the fold, that is modulated by the strongly directional hydrogen bonds and steric hinderance between lateral chains. These latter interactions should obey a fine dynamical network of geometric constraints. We think that the shuffling invariant folding indexes proposed up to now in the literature and re-optimized in the present work are able to capture information related only to the geometry-independent forces, that are globally correlated with a peculiar bias in the amino acid composition of the sequence. Our position, nonetheless, is in line with recent suggestion in the literature \cite{Szilagyi2008}.
To confirm this point, we studied the correlation among folding indexes and the frequencies of amino acids in the protein sequences. To this aim, we used the distinction proposed by Romero $et$ $al.$ in \cite{Romero2001}. They observed that natively unfolded proteins are depleted in \textit{order-promoting residues}: W, C, F, I, Y, V, L; and enriched in \textit{disorder-promoting residues}: M, A, R, Q, S, P, E. We studied the correlation among order- and disorder-promoting amino acid frequencies and mean packing, mean contact energy and $gVSL2$; the results are reported in table \ref{tab:correl}. We observe a high correlation, especially among indexes of fold and frequency of order-promoting amino acids; this confirms that the indexes here investigated are determined by the mere amino acidic composition and not by other more subtle effects, due to a specific order or polarity of the sequences.  \\

\noindent
{\bf Combination of indexes into unanimous and voting scores} \\
We explored the possibility of enhancing the performance of single indexes of fold by combining them into several scoring schemes. We analysed both the unanimous and voting scores by Oldfield $et$ $al.$ \cite{Oldfield2005} and a strictly unanimous score $S_{SU}$ (see Materials and Methods). The results of the predictions are reported in table \ref{tab:Perf_scores}. Comparing table \ref{tab:Perf_idxs} with table \ref{tab:Perf_scores}, we observe that the performance of $S_{U}$ has lower sensitivity with respect to mean packing, mean contact energy and $gVSL2$, whereas $S_{V}$ has higher sensitivity but also a higher level of false predictions. We conclude that $S_{U}$ is less effective than $S_{V}$. On the other hand $S_{V}$ must be used with caution, since the higher level of false predictions may lead to an overestimate of the number of natively unfolded proteins in a given genome. \\

From table \ref{tab:Perf_scores} we observe a significant improvement both in the sensitivity and in the specificity only for the strictly unanimous score $S_{SU}$. On one hand, then, $S_{SU}$ has a higher sensitivity and a lower level of false predictions with respect to all other indexes. On the other hand, $S_{SU}$ leaves unclassified all proteins that mean packing, mean contact energy and $gVSL2$ does not jointly predict in the same class. In our set of 743 folded and 81 natively unfolded proteins, 80 sequences were left unclassified, about 10\% of all proteins; of these 80 unclassified sequences, 15 are natively unfolded, corresponding to 19\% of all natively unfolded proteins in the test set; therefore $S_{SU}$ may have a selectivity bias towards folded proteins. 
It is interesting to note that if when re-evaluated the single folding indexes on the dataset purged by the unclassified proteins by $S_{SU}$, their performance increased and coincided with that of $S_{SU}$. This indicates that proteins unclassified by $S_{SU}$ are mainly false predictions of the single indexes: if we exclude them from the dataset, we reduce the false predictions of the single folding indexes thus increasing their performance. \\

Sequences left unclassified by $S_{SU}$ reasonably have amino acidic composition compatible with both classes; it is assumable that there exists a twilight zone between order and disorder \cite{Szilagyi2008}, and proteins unclassified by $S_{SU}$ belong to that twilight zone. In this zone a single index would be definitely not reliable, haphazardly forcing the assignment of a protein to one or the other class. $S_{SU}$, then, is a conservative reliable index, which refrains from forcing a classification and, positively, useful to select amino acid sequences with a weak folding signature. These left over sequences could be an interesting category per se, or, simply, a group of proteins which challenge the discriminating power of the methods here investigated. \\

\noindent
{\bf Other scoring schemes} \\
We have introduced the score $S_{0}$ to search for a good performance combination score able to take a decision in all cases. It requires consensus among the majority of folding indexes to assign an amino acid sequence to a specific class, so its value could be considered as a quantitative expression of how typically a sequence is assigned to a class or to the other: a higher score means higher consensus among different folding indexes and then a more definite assignment. The performance of $S_{0}$, evaluated on our test set, is reported in table \ref{tab:Perf_scores} and is clearly lower than that of $S_{SU}$; nonetheless the combined use of both indexes can be helpful to reduce the number of unclassified proteins. We applied $S_{0}$ to the 80 proteins left unclassified by $S_{SU}$, and we assigned to a folding class only those with $|S_{0}| > 6$, as shown in the last row of table \ref{tab:Perf_scores} denoted by $S_{SU}/S_{0}$. The combined use of $S_{SU}$ and $S_{0}$ gives a sensitivity of 0.79, a level of false predictions of 0.06 and, of the 80 proteins left unclassified by $S_{SU}$, 46 are still unclassified.  \\

\noindent
{\bf Frequency of disorder in various genomes} \\
In an interesting paper \cite{Ward2004} the classifier DISOPRED2 is used to estimate the disorder frequency in 13 bacterial, 6 archaean and 5 eukaryotic genomes; an average of 4.2\% of eubacterial, 2.0\% of archaean and 33.0\% of eukaryotic proteins are predicted to contain long disordered regions, i.e. segments with at least 30 consecutive disordered amino acids (see table \ref{tab:genomes}). We analysed the same genomes, with the exception of Homo sapiens, by means of the combination scores defined in the above sections (see again table \ref{tab:genomes}). We observe that $S_{0}$ predicts about 5.2\% of eubacterial, 1.7\% of archaean and 22.0\% of eukaryotic proteins as natively unfolded; these percentages are compatible with those predicted using DISOPRED2. 
It is worth noting that the percentage of natively unfolded proteins predicted by $S_{SU}$ are lower than those predicted by $S_{0}$; more precisely, the percentage of natively unfolded proteins predicted by $S_{SU}$ are 3.7\% for Bacteria, 0.8\% for Archaea and 19.3\% for Eukarya. The application of $S_{SU}/S_{0}$, useful to further evaluate sequences left unclassified by $S_{SU}$, gave a quite similar result. 
The results obtained with our scores are correlated with those obtained by means of DISOPRED2 (see figure \ref{fig:correl}), which is a predictor of disordered amino acids that analyses local evolutionary properties of polypeptide chains. Our scores combined different global indicators of folding status, based on the analysis of four basic parameters. The coherence in the predictions obtained through these two different approaches make us confident of the reliability of our predictions. 

It has been suggested that natively unfolded proteins are involved in regulatory and signalling processes inside a cell \cite{Wright1999,Tompa2002,Demchenko2001}. The higher percentage of natively unfolded proteins in Eukarya has been related to: i) the presence of finely regulated degradation pathways that allow disordered proteins to escape recognition processes, strictly based on the structure-function paradigm \cite{Wright1999}; and ii) the necessity of flexible proteins within complex regulatory and signalling networks, typical of eukaryotic organisms \cite{Demchenko2001, Tompa2002}. In fact, it has been observed that, in protein interaction networks, disorder is frequent in the hub proteins \cite{Dunker2005,Heynes2006,Dosztanyi2006}. In figure \ref{fig:genomes} we attempt at establishing a scaling law; on the basis of the genomes here investigated we obtain that the number of natively unfolded proteins, detected by $S_{SU}$, is proportional to the number of proteins in the genome raised to the power $1.95 \pm 0.21$. Further studies are necessary to confirm the validity of this scaling law, possibly relevant for the general biology of genetic code translation but also in the search of allometric relations between frequency of disordered proteins and regulative complexity of the species. \\

\noindent

\vspace{1cm}
\begin{center}
{\bf {CONCLUSIONS}}\\
\end{center}

Let us put in perspective the results obtained in this work. We observed that natively unfolded proteins have, in general, a higher mean contact energy than folded ones; we can relate this property to their difficulty in reaching a stable configuration, corresponding to a relatively low free energy. This explains also their tendency to have a low mean packing, typical of extended conformations, corresponding to minima of the free energy separated by low barriers, of the order of physiological thermal energy scales $k_{B}T_{phys}$. It has been also observed that natively unfolded proteins have a lower mean hydrophobicity and a higher mean net charge \cite{Uversky2000}, and these two parameters have been used to discriminate between the two groups of proteins \cite{Uversky2000,Prilusky2005,Oldfield2005}. As suggested by Uversky \cite{Uversky2000}, natively unfolded proteins do not fold because their hydrophobicity is insufficient, in typical environments, to form the hydrophobic core necessary to nucleate the folding process. It is interesting to observe that mean hydrophobicity and mean contact energy are correlated (Pearson's correlation coefficient equal to -0.74): high hydrophobicity stabilizes the structure and favors the spontaneous search for a minimum free energy configuration. Of course the stabilization of a protein tertiary structure is due not only to hydrophobicity, but also to other forces of different origin (Van der Waals, hydrogen bonding, excluded volume); nonetheless, the strong correlation between hydrophobicity and contact energy supports the idea that contact energy incorporates a strong contribution from hydrophobicity. \\
The invariance of the analysed folding indexes under shuffling of the amino acids in the protein sequences and the correlation between the indexes and the frequency of order-promoting (disorder-promoting) amino acids indicate that these indexes capture information related only to the geometry-independent forces, that are globally related to the amino acid composition of the protein sequences. This fact points to an intrinsic limitation of these approaches in predicting natively unfolded proteins, in line with the observation by Szilagyi $et$ $al.$ \cite{Szilagyi2008}. Proteins with an amino acid composition typical of folded or of natively unfolded sequences are generally correctly identified by single dichotomic predictors of natively unfolded proteins, as indicated by their high performance in dataset purged by unclassified proteins by $S_{SU}$. However, these predictors perform worse in dataset containing proteins with amino acid composition compatible with both folded and natively unfolded sequences, and their performance cannot be significantly enhanced by combining them into score. This last kind of proteins generally are differently classified by single folding indexes, so our strictly unanimous score helps in identify them. Our score $S_{SU}$ therefore can be used in two ways: $i)$ it identifies proteins with strong signature of folding status in their sequence, ideal candidates to be folded or unfolded; $ii)$ it identifies proteins with a weak signature of folding status in their sequence, for which it is haphazarding to say whether they are folded or not, since they are often false predictions of the folding indexes. These proteins are worth to be analysed $per$ $se$ since they may have peculiar structural and functional properties. However, they cannot be analysed with current natively unfolded proteins predictors, but new approaches must be used or developed, possibly considering the order of amino acids in the sequences that influence their interactions among them. 

\begin{center}
{\bf {MATERIALS AND METHODS}}
\end{center}

\noindent
{\bf Datasets}  \\

In this work we used as training set the list of proteins compiled by Prilusky to test FoldIndex \cite{Prilusky2005}, a web-based server aimed at identifying unstructured proteins. It includes 151 folded proteins and 39 proteins reported in the literature as natively unfolded . Folded proteins have a length between 50 and 200 amino acids, they do not contain prosthetic groups or disulphide bridges and their structures have been determined by X-ray cristallography. \\
We compiled our own test set starting from PDBSelect25, version october 2007 \cite{Hobohm1992,Hobohm1994}, that contains 3694 proteins with sequence identity lower than 25\%. To avoid the introduction of poor models we excluded structures with a resolution above 2 {\AA}\hspace{1pt} and an R-factor above 20\%. We obtained a list of 1015 folded proteins. From this list, we extracted a restricted list of 743 fully ordered proteins, that contain less than 5\% of disordered amino acids. We aligned PDB file SEQRES fields with the ATOM fields and the residues that are present in SEQRES but absent in ATOM were considered as disordered. To compile a list of natively unfolded proteins, we started from the DisProt database, version 3.6 \cite{Vucetic2005,Sickmeier2007}. We extracted a list of 81 natively unfolded proteins with at least 95\% of disordered amino acids and sequence identity below 25\%. \\

\noindent
{\bf Mean packing} \\
The mean packing of a protein sequence is the arithmetic mean of the packing values of each amino acid. We used the packing index introduced by Galzitskaya $et$ $ al.$ \cite{Galzitskaya2006a}, based on the number of residues located within a distance of 8 \AA, averaged over a large dataset of structures. We considered a sliding window of length 11 and we assigned its mean packing to the central residue. 

To set the stage we initially computed mean packings on Prilusky's set \cite{Prilusky2005}; we looked for a discriminative threshold as to obtain a sensitivity of at least 0.80 and a level of false predictions as low as possible; we found it at 20.55, getting a sensitivity of 0.82 and a level of false predictions of 0.13. We repeated the experiment with sliding windows of different length, without improvement of the performance. \\

\noindent
{\bf Mean contact energy} \\
We followed the method by Dosztanyi $et$ $al.$ \cite{Dosztanyi2005a}. The contact energy value of an amino acid is a measure of its "contact interaction" with the amino acids located from 2 to 100 positions apart, downward and upward, along the sequence. There are, of course, constraints due to the length of the sequence that should be taken into account in the bookeeping.
The contact energy of amino acid $i$ at position $p$ is given by: \\
\noindent
$e_{i}^{(p)}={\displaystyle\sum_{j=1}^{20} P_{ij} n_{j}^{(p)}}$   \\
where $n_{i}^{(p)}$ is the frequency of amino acid $j$ in a window of length up to 100 around position $p$, taking into account possible limitations on both sides due to the length of the protein. The generic element $P_{ij}$ of the "energy predictor matrix" $P$ expresses the expected contact interaction energy between amino acid $i$ and $j$. \\

Contact energy values are averaged over a window of 21 amino acids and the average is assigned to the central residue at position $p$ in the sequence. Finally, the arithmetic mean of the contact energy values of all the amino acids gives the global mean contact energy of the protein. \\
To discriminate between folded and natively unfolded proteins, we computed mean contact energy of the Prilusky's set \cite{Prilusky2005} and we looked for a threshold, so to get a sensitivity of at least 0.80 and a level of false predictions as low as possible. We found it at -0.37 arbitrary energy unit (a.e.u.), getting a sensitivity of 0.85 and a level of false prediction of 0.14. \\

\noindent
{\bf Index derived from VSL2} \\
VSL2 \cite{Obradovic2005,Peng2006} is a disorder predictor that assigns to each amino acid of a protein sequence the probability that the amino acid is disordered, estimated using a combination of support vector machines. The score from VSL2 is normalized between 0 and 1 and an amino acid is considered disordered if its value is above 0.5. 

We used the arithmetic mean of these disorder scores, evaluated using VSL2B and output windows of length 11, to discriminate folded from unfolded proteins and we call it $gVSL2$ index. We classified a protein as natively unfolded if $gVSL2$ was above 0.5. \\

\noindent
{\bf Combination of two parameters into a single index of fold} \\
We plotted the values of the two parameters on a plane and we looked for discriminative lines. In general there is an overlap region that prevents an exact separation of the two groups of sequences. We identified the overlap region as the narrower vertical band containing points from both groups. For all pairs of points inside the overlap area we traced a line and evaluated its performance in separating the two groups of proteins; among all the discriminative lines with sensitivity above 0.80, we chose that with lowest false predictions. 
If the equation of a discriminative line is: \\
\noindent
$y = a x + b$, \\
\noindent
then the corrisponding scalar index of fold was defined as: \\
\noindent
$I = -sign(\langle{x_{f}}\rangle - \langle{x_{nf}}\rangle) \cdot sign(a) \cdot (y - a x - b)$ \\
\noindent
where $\langle{x_{f}}\rangle$ and $\langle{x_{nf}}\rangle$ are, respectively, the mean values of the index $x$ for folded and natively unfolded proteins. The defined index was positive for folded proteins and negative or 0 for natively unfolded ones. 
If the slope $a$ became very large our code looked for discriminative lines parallel to the ordinate axis and the index was defined as: \\
\noindent
$I = sign(\langle{x_{f}}\rangle - \langle{x_{nf}}\rangle) (x - x_{th})$ \\
\noindent
where $x=x_{th}$ was the optimum discriminative line. \\

\noindent
{\bf Definition of score indexes} \\
We combined mean packing, mean contact energy and $gVSL2$ to obtain score indexes: $S_{U}$, $S_{V}$ and $S_{SU}$, and then $S_0$. $S_{U}$ and $S_{V}$ have been previously proposed by Oldfield $et$ $al.$ \cite{Oldfield2005}. $S_{U}$ is an unanimous score: a protein is classified as natively unfolded if \textit{all the folding indexes} agree on that, otherwise it is classified as folded. $S_{V}$ on the other hand is a voting score: a protein is classified as natively unfolded if \textit{at least one} index assigns it to such a class. We proposed a third combination rule: we classified a protein as folded only if all the indexes predicted it as folded; conversely, we classified a protein as natively unfolded only if all the indexes predicted it as natively unfolded. This rule left a protein unclassified if there is disagreement between at least two indexes. We call this score \textit{strictly unanimous}, $S_{SU}$. 

To obtain $S_{0}$, we increased the number of indexes; we took different pairs of parameters, we plotted their values into planes and obtained an index of fold, as explained in the previous section. We considered all the combinations of the four indexes: Uversky's $HQ$ \cite{Uversky2000}, mean packing, mean contact energy and $gVSL2$ to get 10 new indicators of folding status. We combined them into a global score as follows: if an index predicted a protein as folded, we incremented the score by 1; if the index predicted a protein as unfolded, we decremented the score by 1. We excluded indexes that were unable to discriminate folded from unfolded proteins of the training set with a sensitivity of at least 0.75 and a level of false predictions above 0.15. The score can assume a positive, negative or null value. $S_{0}$ classifies a protein as folded if its value is positive, otherwise it classifies it as natively unfolded. \\

\noindent
{\bf Parameters of performance} \\
To evaluate the performance of the predictors we used very common indicators: \cite{Bordoli2007}:\\
Sensitivity: $S_{n}=\frac{TP}{TP+FN}=\frac{TP}{N_{unfolded}}$,\\
Specificity: $S_{p}=\frac{TN}{TN+FP}=\frac{TN}{N_{folded}}$, \\
False predictions: $f_{p}=1-S{p}=\frac{FP}{TN+FP}$.\\
Where TP stands for True Positive, TN for True Negative, FP for False Positive and FN for False Negative. \\

\begin{center}
{\bf {ACKNOWLEDGMENTS}}
\end{center}

The authors thank Dr. Z. Dosztanyi for sending the IUPred code and the reference \cite{Dosztanyi2007}. \\

\vspace{1cm}
\newpage
\bibliography{ad_ag}
\newpage
\begin{center}
{\bf TABLES}
\end{center}
%


\begin{table}[h]
\begin{center}
\begin{tabular}{|c|c|c|c|}
    \hline 
                             & $S_{n}$ & $S_{p}$ & $f_{p}$ \\ \hline
    $HQ$                     & 0.67    & 0.88    & 0.12    \\ \hline
    $\langle P \rangle$      & 0.74    & 0.93    & 0.07    \\ \hline
    $\langle E_{c} \rangle$  & 0.74    & 0.91    & 0.09    \\ \hline
    $gVSL2$                  & 0.81    & 0.89    & 0.11    \\ \hline
\end{tabular}
\caption{{\bf Performance of single indexes of fold.} Performance of: $HQ$, mean packing, mean contact energy and $gVSL2$ in discriminating natively unfolded proteins among those in test set. $S_{n}$, sensitivity; $S_{p}$, specificity;  $f_{p}$, number of false predictions. See Methods for definitions.  \label{tab:Perf_idxs}}
\end{center}
\end{table}

\begin{table}[h]
\begin{center}
\begin{tabular}{|c|c|c|c|c|c|c|}
    \hline 
                   & $S_{n}$ & $S_{p}$ & $f_{p}$ & $n.c.$ & folded & unfolded \\ \hline
    $S_{U}$        & 0.67    & 0.95    & 0.05    & 0      &  736     &   88   \\ \hline
    $S_{V}$        & 0.85    & 0.87    & 0.13    & 0      &  656     &  168   \\ \hline
    $S_{SU}$       & 0.82    & 0.95    & 0.05    & 80     &  656     &   88   \\ \hline
    $S_{0}$        & 0.73    & 0.93    & 0.07    & 0      &  712     &  112   \\ \hline
    $S_{SU}/S_{0}$ & 0.79    & 0.94    & 0.06    & 46     &  681     &   97   \\ \hline
\end{tabular}
\caption{{\bf Performance of different combination scores.} Performance of the combination scores (see text) on the proteins of the test set. $S_{n}$, sensitivity; $S_{p}$, specificity;  $f_{p}$, number of false predictions; $n.c.$, number of proteins left unclassified. 
\label{tab:Perf_scores}}
\end{center}
\end{table}

\begin{table}
\begin{center}
\begin{threeparttable}
\begin{tabular}{|l|c|c|c|cc|c|}
     \hline
     ORGANISM & N. & DP2\tnote{1} & $S_{0}$ & \multicolumn{2}{|c|}{$S_{SU}$} & $S_{SU}/S_0$ \\
              & proteins & \% $l > 30$ & \% unfolded & \% unfolded & \% n.c. & \% unfolded \\
     \hline
     ARCHAEA                                                                \\ \hline
     A.pernix                    &  1700 &  2.1 &  2.2 &  1.3 &  5.3 &  1.6 \\
     A.fulgidus                  &  2418 &  0.9 &  1.7 &  0.8 &  5.0 &  0.9 \\ 
     Halobacterium sp. \tnote{2} &  2622 &  5.0 & 24.4 & 16.2 & 30.8 & 16.5 \\
     M.jannaschii                &  1768 &  1.0 &  1.1 &  0.2 &  5.4 &  0.5 \\
     P.abyssi                    &  1898 &  1.4 &  1.3 &  0.5 &  5.1 &  0.7 \\
     T.volcanium                 &  1491 &  1.0 &  2.1 &  1.1 &  4.5 &  1.3 \\ \hline
                                 &  9275 &  2.0 &  1.7 &  0.8 &  5.1 &  1.0 \\ \hline
     BACTERIA                                                               \\ \hline
     A.tumefaciens C58           &  5355 &  5.7 &  5.5 &  4.1 &  8.0 &  4.5 \\
     A.aeolicus VF5              &  1558 &  1.9 &  1.5 &  0.5 &  5.9 &  0.7 \\
     C.pneumoniae AR39           &  1085 &  4.8 &  5.8 &  4.1 &  9.0 &  4.7 \\ 
     C.tepidum TLS               &  2247 &  3.3 &  6.2 &  4.7 &  7.7 &  5.3 \\ 
     E.coli K12                  &  4130 &  2.8 &  3.6 &  2.5 &  6.1 &  2.8 \\ 
     H.influenzae Rd             &  1615 &  3.8 &  3.2 &  2.1 &  5.2 &  2.6 \\
     M.tuberculosis H37Rv        &  3989 &  7.0 & 10.1 &  7.4 & 11.6 &  7.9 \\
     N.meningitidis MC58         &  2063 &  4.5 &  6.0 &  4.4 &  8.3 &  4.7 \\
     S.typhi                     &  4756 &  2.7 &  4.2 &  2.9 &  6.8 &  3.2 \\ 
     S. aureus                   &  2618 &  4.5 &  6.6 &  5.5 &  6.9 &  5.9 \\
     Synechocystis PCC 6803      &  3569 &  4.7 &  4.2 &  3.2 &  6.4 &  3.5 \\
     T.maritima                  &  1856 &  1.8 &  2.4 &  1.0 &  5.8 &  1.2 \\
     T.pallidum                  &  1009 &  6.4 &  4.3 &  2.7 &  6.7 &  3.5 \\ \hline
                                 & 35850 &  4.2 &  5.2 &  3.7 &  7.5 &  4.1 \\ \hline
     EUKARYA                                                                \\ \hline
     A.thaliana                  & 31708 & 33.8 & 19.6 & 17.5 & 14.6 & 18.0 \\ 
     C.elegans                   & 22843 & 27.5 & 19.1 & 16.1 & 13.0 & 16.8 \\ 
     D.melanogaster              & 20046 & 36.6 & 29.8 & 26.5 & 14.4 & 27.5 \\
     S.cerevisiae                &  5880 & 31.2 & 19.8 & 17.0 & 14.2 & 17.8 \\ \hline
                                 & 80477 & 33.0 & 22.0 & 19.3 & 14.1 & 20.0 \\ \hline
         
\end{tabular}
\caption{{\bf Frequency of natively unfolded proteins in various genomes\tnote{3}.}
Comparison among the percentage of proteins having disordered segments with more than 30 consecutive amino acids as predicted by DISOPRED2 (DP2) and the percentage of natively unfolded proteins predicted by the scores defined in this work. 
\label{tab:genomes}}
\begin{tablenotes}
  \item [1] From Ward J.J. $et$ $al.$, \textbf{Prediction and functional analysis of native disorder in proteins from the three kingdoms of life}, \textit{J. Mol. Biol.} 2004, \textbf{337}, 635-645
  \item [2] Halobacterium sp. is an outlier, so we did not consider it in the computation of the mean of disordered proteins in the Archaea.
  \item [3] genomes were downloaded from the ftp server of NCBI: ftp://ftp.ncbi.nlm.nih.gov/genomes/
\end{tablenotes}
\end{threeparttable}
\end{center}
\end{table}

\begin{table}
\begin{center}
\begin{tabular}{|c|c|c|}
   \hline
                            & $f_{OP}$ &  $f_{DP}$   \\ \hline
   $HQ$                     &    0.74  &    -0.60    \\ \hline
   $\langle P \rangle$      &    0.91  &    -0.63    \\ \hline
   $\langle E_{c} \rangle$  &   -0.85  &     0.57    \\ \hline
   $gVSL2$                  &   -0.84  &     0.77    \\ \hline
\end{tabular}
\caption{Correlation among fold indexes and frequencies of order- ($f_{OP}$) and disorder-promoting ($f_{DP}$) amino acids \label{tab:correl}}
\end{center}
\end{table}

\newpage

\begin{center}
{\bf {FIGURE CAPTIONS}}
\end{center}

FIGURE 1: {\bf Frequency of natively unfolded proteins in genomes: correlation between combination scores and DISOPRED2.} \\
For each genome considered in table \ref{tab:genomes} the estimate of the average frequency of natively unfolded proteins, estimated with $S_0$, $S_{SU}$ and $S_{SU}/S_0$, are plotted versus the estimate made, using DISOPRED2, by Ward $et$ $al.$ \cite{Ward2004}. The correlation coefficients are: 0.84($S_0$), 0.90($S_{SU}$) and 0.91($S_{SU}/S_0$). \\

FIGURE 2: {\bf Number of predicted natively unfolded proteins vs. total number of proteins in various genomes.} \\
Logarithmic plot of the number of natively unfolded proteins, predicted by $S_{SU}$, vs. the total number of proteins in the genome. The exponent of the power law is: $1.95 \pm 0.21$. \\

\newpage

%
%

%
\begin{figure}
\includegraphics[width=5in,height=4in]{figure1.eps}
\caption{\label{fig:correl}}
\end{figure}

\begin{center}
{\bf Figure 1} Frequency of natively unfolded proteins in genomes: correlation between combination scores and DISOPRED2.
\end{center}

\newpage

\begin{figure}
\includegraphics[width=5in,height=4in]{figure2.eps}
\caption{\label{fig:genomes}}
\end{figure}

\begin{center}
{\bf Figure 2} Number of predicted natively unfolded proteins vs. total number of proteins in various genomes.
\end{center}

\newpage

%

\end{sloppy}
\end{document}